\documentclass[%
 reprint,
superscriptaddress,
 amsmath,amssymb,
 aps,
]{revtex4-1}

\usepackage{graphicx}
\usepackage{dcolumn}
\usepackage{xcolor}
\usepackage{bm}
\usepackage{hyperref}

\newcommand{\ket}[1]{|#1\rangle}

\usepackage[normalem]{ulem}

\begin{document}
\preprint{APS/123-QED}
\title{Nuclear spin gyroscope based on the NV center in diamond}
\author{Vladimir V. Soshenko}
\affiliation{P. N. Lebedev Physical Institute, 53 Leninskij Prospekt, Moscow, 119991, Russia}
\affiliation{LLC Sensor Spin Technologies, The Territory of Skolkovo Innovation Center, Str. Nobel b.7, Moscow, 143026, Russia}
\author{Stepan V. Bolshedvorskii}
\affiliation{P. N. Lebedev Physical Institute, 53 Leninskij Prospekt, Moscow, 119991, Russia}
\affiliation{LLC Sensor Spin Technologies, The Territory of Skolkovo Innovation Center, Str. Nobel b.7, Moscow, 143026, Russia}
\affiliation{Moscow Institute of Physics and Technology, 9 Institutskiy per., Dolgoprudny, Moscow Region, 141700, Russia}%
\author{Olga Rubinas}
\affiliation{P. N. Lebedev Physical Institute, 53 Leninskij Prospekt, Moscow, 119991, Russia}
\affiliation{LLC Sensor Spin Technologies, The Territory of Skolkovo Innovation Center, Str. Nobel b.7, Moscow, 143026, Russia}
\affiliation{Moscow Institute of Physics and Technology, 9 Institutskiy per., Dolgoprudny, Moscow Region, 141700, Russia}%
\author{Vadim N. Sorokin}
\affiliation{P. N. Lebedev Physical Institute, 53 Leninskij Prospekt, Moscow, 119991, Russia}
\author{Andrey N. Smolyaninov}
\affiliation{LLC Sensor Spin Technologies, The Territory of Skolkovo Innovation Center, Str. Nobel b.7, Moscow, 143026, Russia}
\author{Vadim V. Vorobyov}%
\affiliation{P. N. Lebedev Physical Institute, 53 Leninskij Prospekt, Moscow, 119991, Russia}
\affiliation{3rd Institut of Physics, IQST and Centre for Applied Quantum Technologies, University of Stuttgart, Pfaffenwaldring 57, 70569 Stuttgart, Germany}%
\author{Alexey V. Akimov}
\affiliation{P. N. Lebedev Physical Institute, 53 Leninskij Prospekt, Moscow, 119991, Russia}
\affiliation{Texas A\&M University, 4242 TAMU, College Station, TX 77843, USA}
\email{akimov@physics.tamu.edu}

\begin{abstract}
A rotation sensor is one of the key elements of inertial navigation systems and compliments most cellphone sensor sets used for various applications. Currently, inexpensive and efficient solutions are mechanoelectronic devices, which nevertheless lack long-term stability. Realization of rotation sensors based on spins of fundamental particles may become a drift-free alternative to such devices. Here, we carry out a proof-of-concept experiment, demonstrating rotation measurements on a rotating setup utilizing nuclear spins of an ensemble of NV centers as a sensing element with no stationary reference. The measurement is verified by a commercially available MEMS gyroscope.
\end{abstract}

\maketitle
\section*{Introduction}

With the increasing interest in unmanned and autonomous vehicles, inertial navigation becomes of pivotal importance for steering in areas lacking global navigation system (GPS) signals, for example, inside buildings or tunnels and under water or ground \cite{Inertial_sensors2018, Vanegas2016, Khattab2015}.
The precision of such inertial navigation is strongly dependent on the number of inertial sensors, among which is the gyroscope. The records for precision and bias stability of industrially available gyroscopes are traditionally held by ring-laser gyroscopes and fiber-optic gyroscopes, the ideas of which were developed by the end of the 20th century \cite{Etrich1992, Ferrar1989ProgressIF, Bergh1981AllSM, Lefvre2014TheFG, Lefvre1989TheFG}.
Based on the Sagnac effect, their precision is proportional to the surface enclosed by the optical light path \cite{Arditty81,Sagnac}.
While there has been great progress in miniaturization of this type of device \cite{Lai2020}, fundamental limits, related to the size and sensitivity of the device, are still hard to overcome.
On the other hand, much less precise microelectromechanical system (MEMS) gyroscopes are widely used for mass production in consumer electronics.
While the bias stability of these devices is often not sufficient for robust long-term inertial navigation, these devices have excellent power consumption characteristics, chip-scale dimensions and low prices \cite{Passaro2017GyroscopeTA}.
Despite enormous progress in improving the bias stability and precision of these devices, there is still a considerable gap between compact and precise devices.

As an attempt to bridge the gap, redesigning laser-based gyroscopes by trying to use chip-scale high quality factor cavities \cite{Yang2018,Lai2020} is currently at the stage of implementing prototypes capable of measuring the earth rotation rate. Another competing approach is to develop hyperpolarized noble gas nuclear spin gyroscopes \cite{Woodman1987,Kornack2005, Walker2016,Vershovskii2018,Zhang2020}. These gyroscopes can reach the precision of a state-of-the-art ring-laser gyroscope with a much smaller form factor of the sensing element. A possible approach to miniaturizing gyroscopes while maintaining high precision is to use solid-state spin systems. The nitrogen-vacancy (NV) center in diamond has demonstrated excellent properties as a solid-state spin system. It possesses optically detectable electron spin, with a long coherence time at room temperature and means for optical polarization of the spin state. Several key applications of NV centers have been demonstrated, including precise measurement of constant and oscillating magnetic fields \cite{Wolf2015,Schloss2018}, nuclear magnetic resonance spectroscopy with chemical resolution \cite{Aslam2017, Glenn2018}, nanoscale electric field sensing and temperature sensing inside living cells \cite{Dolde2011,Maletinsky2012, Kucsko2013b}.

In 2012, it was proposed \cite{Ledbetter2012b, Ajoy2012d} to use an NV center ensemble as a solid-state nuclear spin gyroscope. Later in \cite{Woodeaar7691,Wood2020}, quantum sensing of a rapid rotation (200 kHz) with a single NV electron spin was demonstrated. The potential for utilization of the nuclear spin was demonstrated recently \cite{Jaskula2019}. However, the realization of a rotating sensor utilizing a nuclear spin ensemble in diamond remains a challenging task.
In this letter, we report the first proof-of-principle direct gyroscopic measurement of a sub-Hz rotation using a hyperpolarized $^{14}N$ nuclear spin $I=1$ ensemble in the solid state. Though the initial ideas were based on geometric Berry phase detection, here, we utilize dynamic phase acquisition as a result of a pseudomagnetic field induced by the rotation \cite{Tycko1987}. We use an ensemble of NV centers in diamond to hyperpolarize the nuclear spin using recursive nuclear spin initialization \cite{Pagliero2014c,Chakraborty2017,Smeltzer2009}, followed by double quantum Ramsey spectroscopy \cite{Mamin2014} on a hyperpolarized nuclear spin qutrit and nuclear spin readout via the optically detected electron spin of NV centers. We perform cross-sensor feedback \cite{Jaskula2019} and subtract systematical shifts using a cothermometer and a comagnetometer based on the same ensemble of NV centers in a diamond sample. We calibrate our sensor on a rotation platform using a commercially available MEMS device.

\begin{figure*}
\begin{center}
\includegraphics[width=\textwidth]{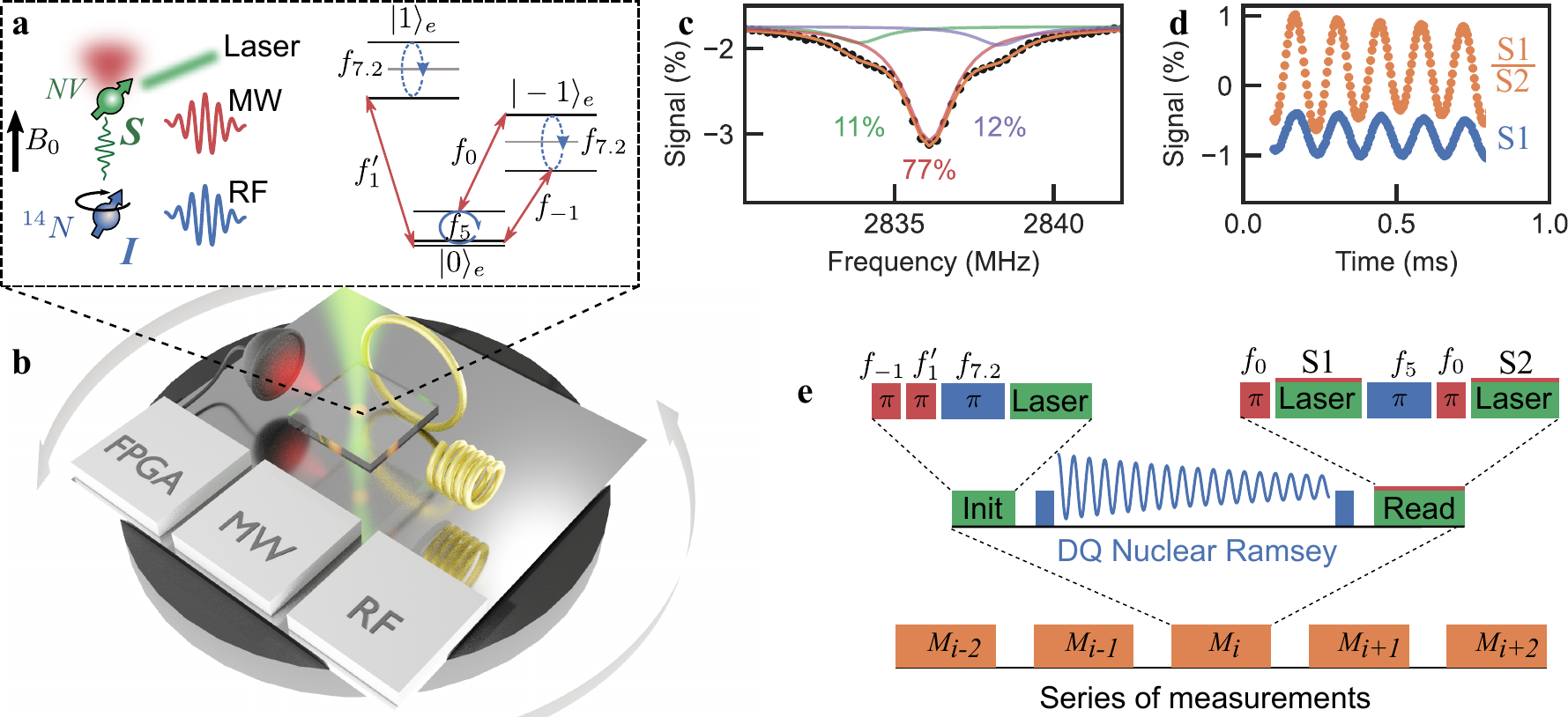}
\caption{Scheme of the experiment. \textbf{a)} NV center in diamond associated with the nuclear spin of $^{14}N$. It is addressed with laser light, microwave (MW) and radiofrequency (RF) control. The fluorescence of the NV center is collected using a photodiode. \textbf{b)} The whole experimental setup with ensembles of NV centers inside a diamond plate is positioned on a rotation stage to perform measurement of a calibrated rotation at various speeds. \textbf{c)} ODMR spectrum of the NV center ensemble with initialized nuclear spin. The nuclear spin is initialized before each frequency is measured. \textbf{d)} Readout of the Ramsey nuclear spin. Orange is the optimized referenced readout, and blue is a simple readout with only a single pi pulse and a single laser pulse. \textbf{e)} Measurement sequence for a gyroscopic measurement. Each measurement consists of an initialization step and nuclear spin free evolution in a scheme of double quantum interference}
\label{fig1}
\end{center}
\end{figure*}

\section*{Results}

All NV centers in diamond contain nitrogen nuclear spin, which in natural abundance is 99.6 \% $^{14}N$ with spin 1. Since nitrogen is part of the NV color center, it has a well-determined interaction with the NV electron spin (Fig. \ref{fig1}a).
Therefore, compared to the more developed nuclear spin of carbon-13, which is often believed to have better coherence and storage properties \cite{Maurer2012}, the nuclear spin embedded into NV center nitrogen is suitable for ensemble measurements.
Namely, using known techniques, one can polarize and readout nitrogen spin via the electron spin of an NV center spin \cite{Smeltzer2009}.
The electron spin structure of the NV center ground state (GS) can be seen as a V-type scheme (see \ref{fig1} a) with the $m_s=0$ subdomain as the lower state and $m_s=-1$ and $m_s=1$ as the upper states. Nevertheless, each of these states has hyperfine splitting due to the interaction with the nitrogen nuclear spin, which can be described by the following Hamiltonian:

\begin{equation}
H = D S_z^2 + \gamma_e  S_z B_z + S \textbf{A} I + Q I_z^2 + \gamma_n B_z I_z
\end{equation}
where $D$ and $Q$ are the zero-field splittings of the electron and nuclear spins, $\gamma_e$ and $\gamma_n$ are the gyromagnetic ratios of the electron and nuclear spin, $S$ and $I$ are spin operators for the spin $1$ system, and $A$ is the hyperfine tensor. Due to the different hyperfine splittings in electron spin subdomains, microwave (MW) transitions allow flipping of electron spin selectively on the nuclear spin state \cite{Smeltzer2009}.
Similarly, radiofrequency (RF) transitions can selectively flip nuclear spin.
However, at small magnetic fields of $\approx 10$ Gauss, the transition frequencies in $m_s=0$ between $m_i=0$ and $m_i=\pm 1$ are not resolved within the natural width of the transition and can be addressed with a single frequency pulse at $Q/2\pi \hbar \approx 5 \, \mathrm{MHz}$ ($f_5$).
Additionally, for $m_s=1$ and $m_s=-1$, transitions from $m_i=0$ to $m_i=1$ and $m_i=-1$, respectively, have similar frequencies of approximately $(Q-A_{||})/2\pi \hbar \approx 7.2 \, \mathrm{MHz}$ ($f_{7.2}$) and can be addressed with a single frequency pulse.
We use this to perform effective control of the system using only two RF frequencies.

Our gyroscopic measurement consists of three main elements: nuclear spin polarization, free precession, and nuclear spin readout.
The polarization of the nuclear ensemble into the $m_i=0$ spin state is performed using a recursive transfer of population \cite{Pagliero2014c,Chakraborty2017,Smeltzer2009} as depicted in Fig. \ref{fig1} a, c.
The first step of the sequence is the transfer of the population from the $m_s=0$ electron spin state to $m_s=1$ and $m_s=-1$ conditioned on nuclear spin $m_i=-1$ and $m_i=1$, respectively, using spectrally narrow enough microwave (MW) $\pi$ pulses.
The next step is to apply a single spectrally broad radiofrequency (RF) $\pi$ pulse that transfers both populations simultaneously to the $m_i=0$ sublevel of the $m_s=1$ and $m_s=-1$ states.
Finally, a green laser pulse is applied to transfer the population to the $m_s=0$ electron spin level by means of optical pumping, in which case the NV center is known to mostly populate the $m_i=0$, $m_s=0$ state. This procedure is repeated 4 times to achieve the maximum population in the $m_i=0$ state of $77 \pm 1$\% (see Fig. \ref{fig1}e).
The nuclear spin readout is performed with a single iteration of the reading sequence depicted in Fig. \ref{fig1}e. The nuclear spin state is transformed into a fluorescence contrast signal using a selective microwave $\pi$ pulse applied to the central $m_i=0$ peak of the NV optically detected magnetic resonance (ODMR) triplet (see Fig. \ref{fig1}c),
followed by optical readout of the electron spin referenced to the same measurement with an added RF $\pi$ pulse of frequency $f_5$.
This referencing technique doubles the fluorescence contrast and subtracts the intensity noise of the laser and MW power fluctuations (see Fig. \ref{fig1}d).
\begin{figure}
\begin{center}
\includegraphics[]{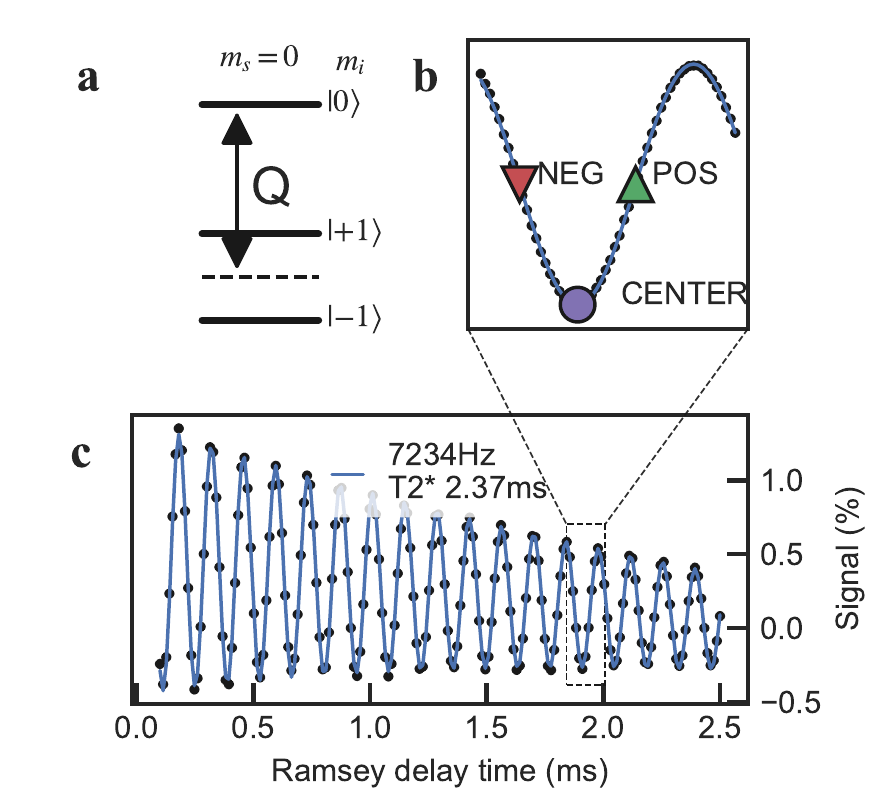}
\caption{Rotation signal. \textbf{a)} DQ Ramsey transition scheme for $m_s = 0$. b) Zoomed Ramsey fringe at working point of the gyroscope. NEG, POS represents negative and positive slope working points for alternating Ramsey protocol c) Double quantum Ramsey spectroscopy of $^{14}N$ nuclear spin. The oscillation matches the nuclear Zeeman splitting, $T_2^*=2.37$ ms}
\label{fig2}
\end{center}
\end{figure}
The rotation of the setup around the main quantization axis of the NV center ensemble (111 axis of the diamond) is analogous to the introduction of a pseudomagnetic field \cite{Tycko1987}, where the nuclear spin behavior is described by the Hamiltonian:
\begin{equation}
H_0 = Q I_z^2 + \gamma_n B_z I_z + \Omega I_z
\end{equation}
To acquire a rotation signal, we monitor the free precession of the $^{14}N$ nuclear spin ensemble in the electron spin $m_s = 0$ subdomain of the NV center ground state (Fig. \ref{fig1}e). Free precession of the nuclear spin after initialization is started with a broadband radiofrequency $\pi$ pulse of frequency $Q/2\pi\hbar$ ($f_5$) (Fig. \ref{fig2}a), which brings the nuclear spin into the "bright" superposition state $\ket{b}=(\ket{1}+\ket{-1})/\sqrt{2}$.
During the free evolution under the Hamiltonian $H_0$, this state acquires a phase $\phi = 2 (\gamma_n B_z+\Omega) \tau$ due to the energy splitting between the $m_i = \pm 1$ sublevels. The state after the free evolution can be expressed as
$(e^{i\phi}\ket{1}+e^{-i\phi}\ket{-1})/\sqrt{2}=(\cos\phi \ket{b} + \sin \phi \ket{d})/\sqrt{2}$, where $\ket{d}=(\ket{1}+\ket{-1})/\sqrt{2}$ denotes the "dark" state. The last $\pi$ pulse converts the bright state back to $\ket{0}$, while the dark state remains unchanged. The population in the $\ket{0}$ nuclear spin state is then transformed into a measurable fluorescence contrast. The experimental double quantum Ramsey precession of $^{14}N$ is shown in Fig. \ref{fig2}b,c.
It is fitted with a decaying sinusoidal curve with $T_2^{*} = 2.37 $ ms. By fixing a working point at $\tau = 2$ ms, we recalculate the rotation signal from the measured fluorescence (Fig. \ref{fig2}b):
\begin{equation}
\Delta \Omega = \frac{1}{a(t_p + t_n)}(S_p(\Delta \Omega)-S_N(\Delta \Omega))
\end{equation}
where $a = 0.5 (S_{max}-S_{min})$ is the amplitude of Ramsey fringes at the selected interrogation time (see SI) and $t_p$ and $t_n$ are the times of positive and negative slope, respectively, which we use to remove the uncertainties due to the low frequency noise related to the initialization fidelity of the nuclear spin and the fidelities of the $\pi$ pulses.

\begin{figure}
\begin{center}
\includegraphics[width=\columnwidth]{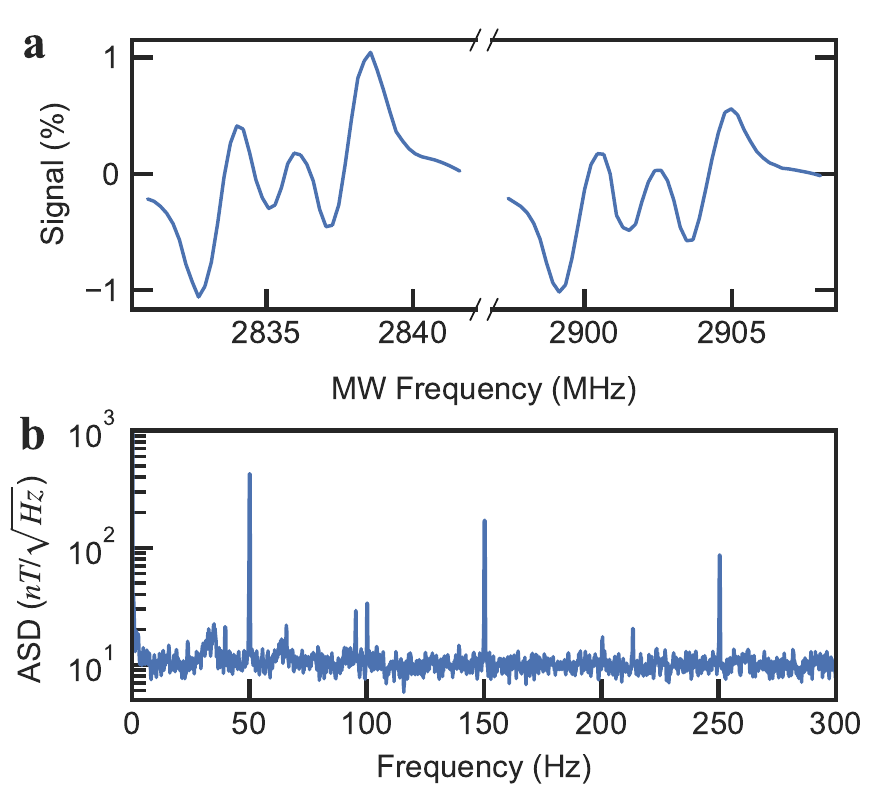}
\caption{Magnetometer and thermometer performance. \textbf{a)} ESR dispersion contour when scanning the $m_s = -1$ and $m_s = 1$ electron spin transitions. \textbf{b)} Amplitude spectral density (ASD) of a free running magnetometer}
\label{fig3}
\end{center}
\end{figure}

\begin{figure*}[t]
\begin{center}
\includegraphics[width=\textwidth]{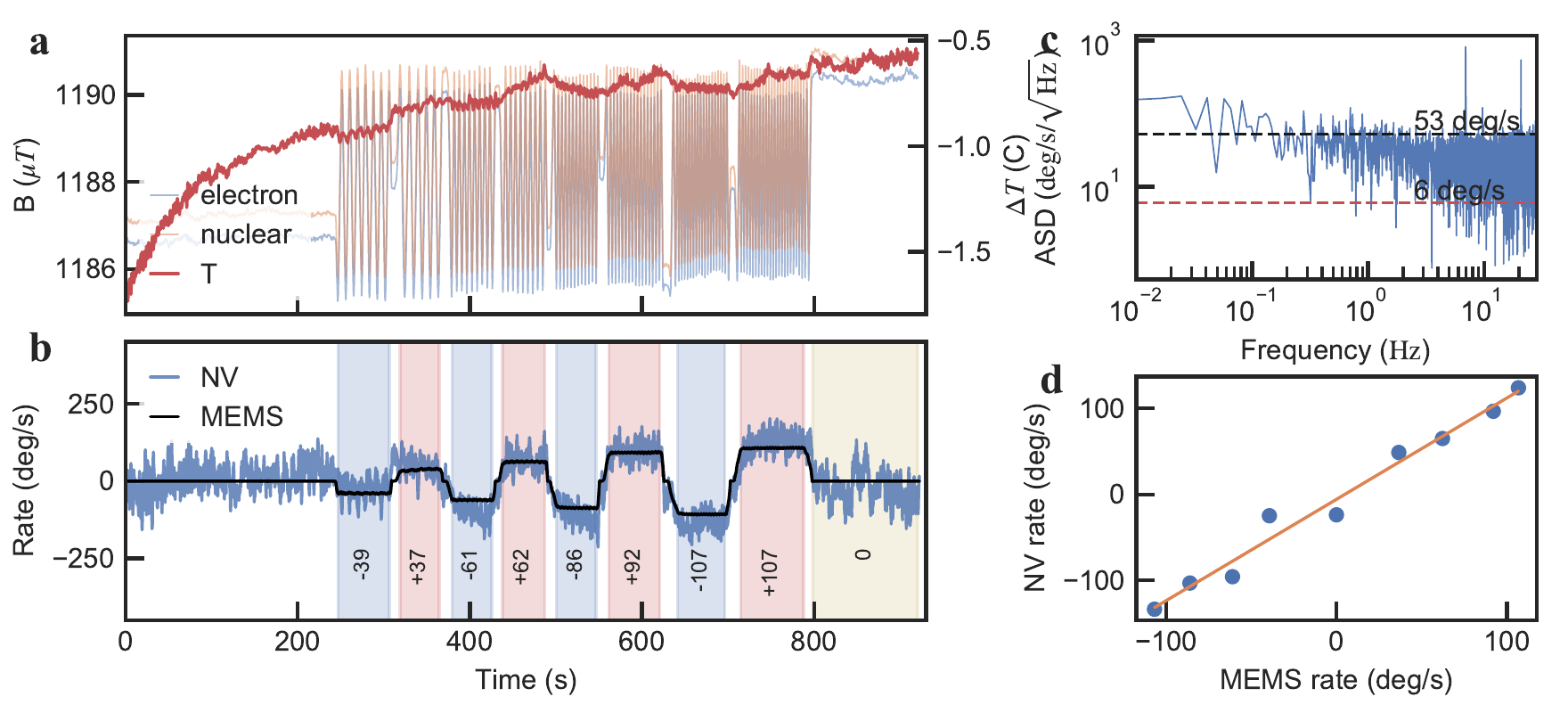}
\caption{Calibration measurement on a rotation stage. \textbf{a)} Sensing of rotation parameters on the rotation table. Data of a comagnetometer and a cothermometer in various rotation regimes. \textbf{b)} Sample MEMS gyro and rotation signal from the NV gyro. Scaling linearity of the sensor output with applied rotation. The black solid curve is the sampling value of a commercial MEMS gyroscope, and the blue solid line is the NV gyro sensor output. \textbf{c)} Amplitude spectral density (ASD) of noise of the sensor output estimated in the last part of the measurement run, where the rotation speed was zero. The red curve is the estimation of the ASD based on the photon shot noise of the detected fluorescence. The black dashed curve is the noise floor of the NV-based sensor. \textbf{d)} NV gyroscope signal output at various rotation rates of the moving stage. The error bars are within the marker size and estimated as the standard variance divided by the square root of the number of outputs. We note a linear character of the rotation signal dependence as a MEMS rotation signal output with a slope of $1\pm$}
\label{fig4}
\end{center}
\end{figure*}

Due to the laser irradiation and electrical current in nearby antennas, the diamond sample unavoidably heats up. Temperature shifts affect both electron \cite{Acosta2010} and nuclear hyperfine terms in the Hamiltonian as in \cite{Soshenko2018, Jarmola2020, Barson2019}.
In addition, the position of the ODMR resonances is also sensitive to stray external magnetic fields, which may change over time.
Thus, the sensor requires compensation of these temperature- and magnetic field-induced shifts.
To realize this, the temperature and magnetic field were measured using the same NV ensemble between the gyroscopic measurements similar to in \cite{Jaskula2019}.
In our case, we implemented digital frequency modulation, thus measuring both  transitions (Fig. \ref{fig3}a). The results of this measurement were used to extract magnetic field and temperature values, which were then used in the feedback loops.
The sensitivity of such a realized comagnetometer has a noise floor of $10 \; \mathrm{nT/\sqrt{Hz}}$ (see Fig. \ref{fig3}b), which corresponds to a rotation measurement noise floor of $11\; \mathrm{deg/s/\sqrt{Hz}}$ (see SI). The magnetic field and temperature measurements were performed between two sequential gyroscopic measurements, and their results were used as feedback to subtract the corresponding systematic noise in the rotation sensor (see SI). Having at hand information about both ESR transitions, we also realized a cothermometer and compensated for the temperature-related shifts (see SI).

Demonstration of a proof-of-principle gyroscope sensor is a challenging task since the whole experimental setup should be rotated. Vibrational mechanical noise could reduce the optical signal-to-noise ratio, thus causing degradation of the performance of the sensor. To avoid difficulties with rotational electrical and optical joints, we employed full rotation of the optical setup and all devices in use. To this end, we assembled an experimental setup that can autonomously operate on a rotation stage (see. Fig. \ref{fig1} b).
It is equipped with a battery power supply, wireless communication protocols, field programmable gate arrays (FPGA) a control board, a laptop and all microwave (MW), radiofrequency (RF) and optical equipment required for operation (see SI).

To calibrate our sensor, a series of continuous rotations with various rotation speeds in both directions were performed (see Fig. \ref{fig4}a,b).
We obtained a series of measurements from the comagnetometer, cothermometer, gyroscope, and calibration MEMS gyroscope, which are depicted in Fig. \ref{fig4}a,b.
As clearly seen, the results of the calibration MEMS gyroscope clearly correlate with the outputs of our NV-based gyroscope sensor.
The obtained rotation signal is linearly proportional to the real rotation speed, as depicted in Fig. \ref{fig4}d.
The noise floor of our NV gyroscope was measured to be $52  \; \mathrm{deg/s/\sqrt{Hz}}$ (see Fig. \ref{fig4}c), which was estimated as the noise floor of the power spectral density (PSD) of the time trace of the gyro output.

Thus, in this paper, we demonstrated direct measurement of rotation using nuclear spins of NV centers. The measurement does not utilize any inertial reference or other nonrotation reference. While the precision of the gyro at this stage is lower than that of the MEMS device, this result, together with a previous demonstration of long-term stability on a nonrotating device \cite{Jaskula2019}, paves the way for a solid-state diamond-based gyroscope with superior bias stability.

\section*{Conclusions}
We showed a direct signal of rotation measured with a nuclear spin gyroscope realized based on an ensemble of NV centers in diamond. By simultaneously measuring the magnetic field and temperature, we subtracted the systematic shifts from the rotation signal and obtained a linear calibration curve of the rotation signal obtained with our NV gyroscope. The results prove the possibility of measuring rotation using nuclear spins of an ensemble of NV centers in diamond and pave the way for low drift compact solid-state gyroscopes.


\section*{Acknowledgments}
This work was supported by the noncommercial organization "Foundation for Development of the New Technologies Development and Commercialization Centre" ("Skolkovo Foundation") grant \# - G8/17 23.03.2017.
This work was also supported by Russian Science Foundation Grant \#16-19-10367 (experimental measurement of the temperature shift).
\bibliographystyle{ieeetr}
\bibliography{references}
\end{document}


\preprint{APS/123-QED}
\title{Supplementary materials for Nuclear spin gyroscope \\ based on the NV center in diamond}
\author{Vladimir V. Soshenko}
\affiliation{P. N. Lebedev Physical Institute, 53 Leninskij Prospekt, Moscow, 119991, Russia}
\affiliation{LLC Sensor Spin Technologies, The Territory of Skolkovo Innovation Center, Str. Nobel b.7, Moscow, 143026, Russia}
\author{Stepan V. Bolshedvorskii}
\affiliation{P. N. Lebedev Physical Institute, 53 Leninskij Prospekt, Moscow, 119991, Russia}
\affiliation{LLC Sensor Spin Technologies, The Territory of Skolkovo Innovation Center, Str. Nobel b.7, Moscow, 143026, Russia}
\affiliation{Moscow Institute of Physics and Technology, 9 Institutskiy per., Dolgoprudny, Moscow Region, 141700, Russia}%
\author{Olga Rubinas}
\affiliation{P. N. Lebedev Physical Institute, 53 Leninskij Prospekt, Moscow, 119991, Russia}
\affiliation{LLC Sensor Spin Technologies, The Territory of Skolkovo Innovation Center, Str. Nobel b.7, Moscow, 143026, Russia}
\affiliation{Moscow Institute of Physics and Technology, 9 Institutskiy per., Dolgoprudny, Moscow Region, 141700, Russia}%
\author{Vadim N. Sorokin}
\affiliation{P. N. Lebedev Physical Institute, 53 Leninskij Prospekt, Moscow, 119991, Russia}
\author{Andrey N. Smolyaninov}
\affiliation{LLC Sensor Spin Technologies, The Territory of Skolkovo Innovation Center, Str. Nobel b.7, Moscow, 143026, Russia}
\author{Vadim V. Vorobyov}%
\affiliation{P. N. Lebedev Physical Institute, 53 Leninskij Prospekt, Moscow, 119991, Russia}
\affiliation{3rd Institut of Physics, IQST and Centre for Applied Quantum Technologies, University of Stuttgart, Pfaffenwaldring 57, 70569 Stuttgart, Germany}%
\author{Alexey V. Akimov}
\affiliation{P. N. Lebedev Physical Institute, 53 Leninskij Prospekt, Moscow, 119991, Russia}
\affiliation{Texas A\&M University, 4242 TAMU, College Station, TX 77843, USA}
\email{akimov@physics.tamu.edu}

\maketitle
\tableofcontents
\section*{Comagnetometer, Cothermometer}
To implement the comagnetometer, we used a continuous probing scheme probing both the $m_s=-1$ and $m_s=+1$ peaks, where the microwave and laser were always switched on. The microwave frequencies were modulated with a square wave signal. The modulation frequencies for the $m_s=-1$ and $m_s=+1$ peaks were \SI{2}{kHz} and \SI{4}{kHz}, respectively. The modulation span was \SI{1}{MHz}. The magnetometer acquisition time was \SI{3}{ms} (6 full measurement cycles).
The microwave power for the magnetometer was attenuated to the nominal microwave power used in the gyroscope part of the experimental sequence by passing the MW signal through \SI{20}{dB} attenuators (Minicircuits) via an additional switch in the MW signal path (Fig \ref{fig2_SI}).

\section*{Temperature compensation}
\begin{figure*}
\begin{center}
\includegraphics[]{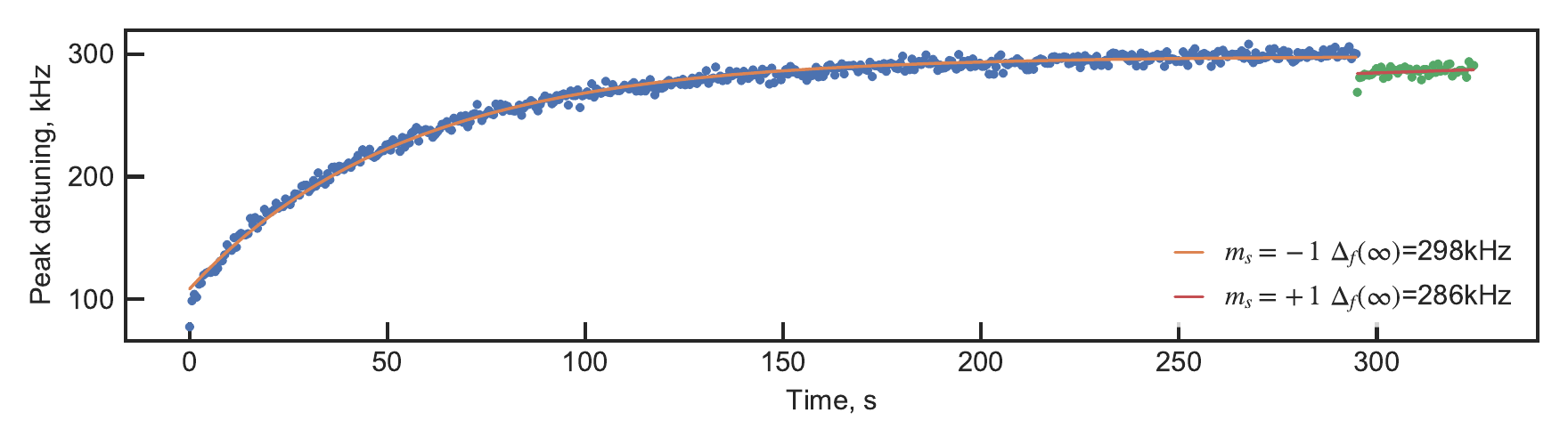}
\caption{Thermalization of the diamond after gyroscope startup. The peak detuning is the difference between the electronic resonance positions measured during gyroscopic measurement and the ODMR experiment}
\label{fig1_SI}
\end{center}
\end{figure*}

\begin{figure*}
\begin{center}
\includegraphics[]{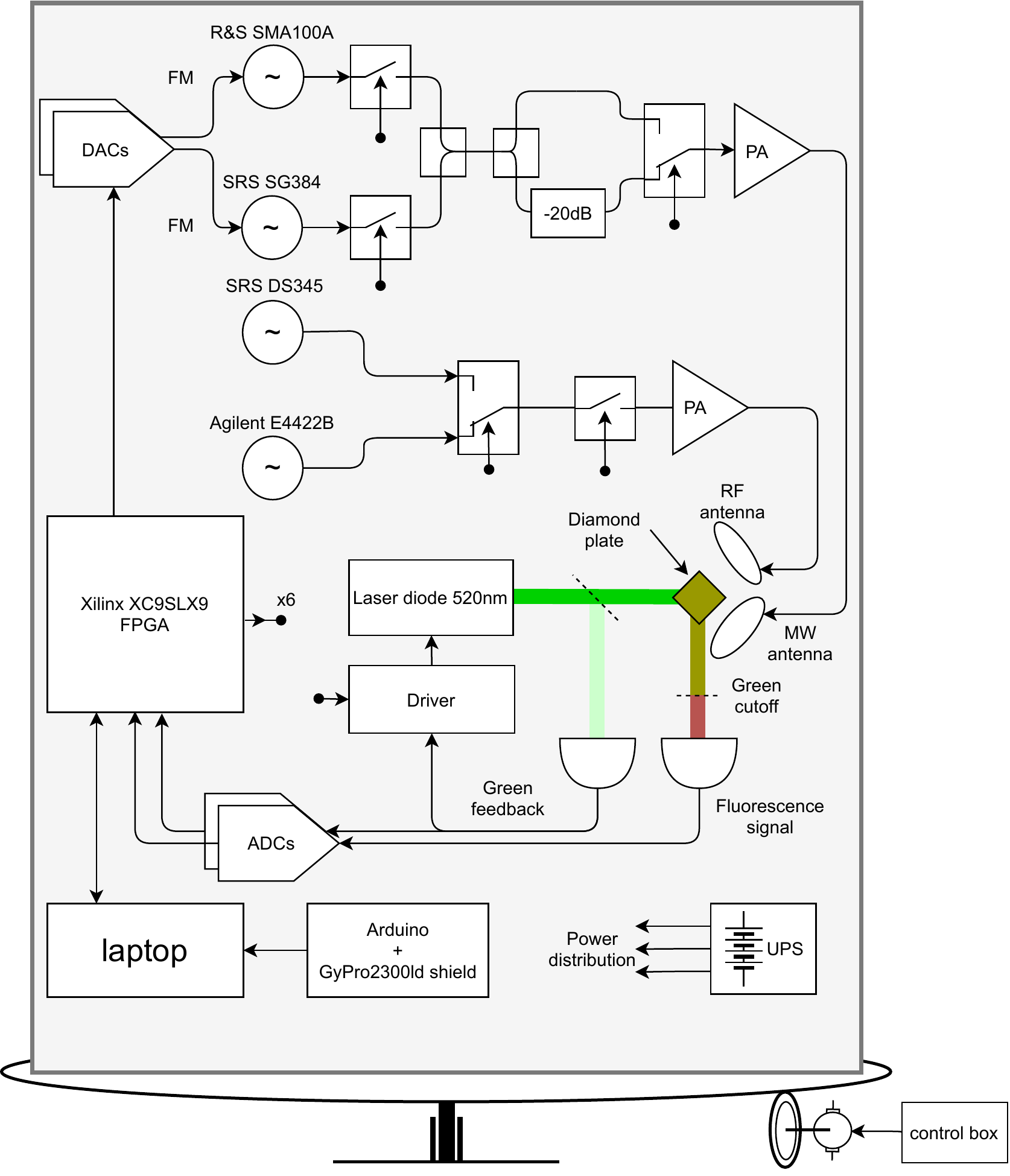}
\caption{Scheme of the setup used for gyroscopic measurements. DAC – digital-to-analog converter, FM – frequency modulation, PA – power amplifier, FPGA – field gate programmable array, RF antenna – radiofrequency antenna for nuclear transitions, MW – microwave antenna, ADC – analog-to-digital converter, UPS – uninterruptible power supply. Black dots represent digital signals connected from FPGA digital outputs to the input of the switches.}
\label{fig2_SI}
\end{center}
\end{figure*}

As an initial approximation for the rotation experiment, we defined the electron resonance frequencies from the ODMR experimental sequence. 
However, the diamond temperature is influenced by heat from the pumping light and radiofrequency signals. 
Thus, changes in the experimental sequence, i.e., between the ODMR experiment and the rotation measurement, will change the diamond temperature and lead to unwanted shifts in the electron resonances. 
To overcome this, we performed rotation measurements (depicted in Fig. 1e of the main text) with conventional readout of electronic spin and a microwave frequency sweep from run to run. This allowed us to reveal the unshifted electron resonance positions and the setup heat-up time, as depicted in Fig. \ref{fig1_SI} With an exponential time constant of \SI{54}{s}, both electron resonances shifted by approximately \SI{+300}{kHz}, caused by a temperature shift of approximately \SI{-4}{K}.
When the rotation measurement started, the first 200 seconds of data were omitted to allow diamond thermalization to finish.

\section*{Equivalent magnetic field for the rotation rate}
The gyro signal includes both the rotation signal and magnetic field as written below:
\begin{equation}
\Delta \Omega = 2 (\Omega_{Rot} + \gamma_{n} \Delta B_z)
\end{equation}
where $\gamma_{n}$ is the gyromagnetic ratio of $^{14}N$, and the factor of 2 is because the Ramsey phase is accumulated for the superposition of states with $\Delta m_I = 2$. To exclude the magnetic field part of the signal, a comagnetometer is utilized. The comagnetometer operates in interleaved mode, as its sensitive part is the same diamond plate. The noise in the magnetometer output then relates to the gyro output through this formula with a proper unit conversion:
\begin{equation}
\delta \Omega [ \mathrm{^o/s} ] = \delta B [ \mathrm{T} ] \gamma_{n}[ \mathrm{Hz/T}] \cdot 360 [ \mathrm{^o}  ]
\end{equation}

\section*{Recalculation of the rotation signal from fluorescence}
The working point (that is, the right free precession time) for gyro operation is selected by sweeping the free precession time near 2 ms (optimal for measurement due to decoherence) and selecting the time point with the maximum derivative (that is, $1944 \;\mathrm{\mu s}$). The Ramsey signal can be described as:
\begin{equation}
S_R (\omega, t)= a \cos \left( \omega t + \right) + b
\label{eq:ramsey}
\end{equation}
where $\Omega_0$ is the Ramsey beating frequency at the current magnetic field ($7200 \; \mathrm{Hz}$), $a$ is the  oscillation amplitude, $b$ is a fluorescence offset, and $T_0$  is the working time (i.e., time of rotation integration). To subtract the offset noise coming from instabilities of the fluorescence level and nuclear spin polarization level, we utilize the fluorescence from two consecutive measurements with alternating free precession times adjusted to be on a negative and positive slope of the Ramsey fringes as described below.
$t_n$ and $t_p$ are selected so that $\Omega_0 t_n + \phi_0 = (2N + \frac{1}{2})\pi$ and $\Omega_0 t_p + \phi_0 = (2N + \frac{3}{2})\pi$. If we add $\Delta \Omega$ to the Ramsey beating frequency and put it into equation \ref{eq:ramsey}, then we obtain:
\begin{equation}
\begin{split}
&S_n (\Delta \Omega) = S_R (\Omega_0 + \Delta \Omega, t_n) = \\ & a \cos((\Omega_0 + \Delta \Omega) t_n + \phi_0) + b = - a \sin(\Delta \Omega t_n) + b
\end{split}
\end{equation}

\begin{equation}
\begin{split}
&S_p (\Delta \Omega) = S_R (\Omega_0 + \Delta \Omega, t_p) = \\ & a \cos((\Omega_0 + \Delta \Omega) t_n + \phi_0) + b = a \sin(\Delta \Omega t_n) + b
\end{split}
\end{equation}
By subtracting $S_n$ from $S_p$, we obtain:
\begin{equation}
\begin{split}
&S_p(\Delta \Omega) - S_n(\Delta \Omega) =  a \left( \sin \Delta \Omega t_p + \sin \Delta \Omega t_n \right) \\ &\approx a \Delta \Omega (t_p + t_n)
\end{split}
\end{equation}
Finally, for $\Delta \Omega$:
\begin{equation}
\Delta \Omega = \frac{1}{a (t_p + t_n)} (S_p(\Delta \Omega)- S_n(\Delta \Omega))
\end{equation}

\section*{Setup details}
The main signal control and conditioning unit is a home-built control device with a Spartan-6 FPGA circuit, providing analog signal digitization, analog signal output for generator modulation and digital pulse generation with \SI{10}{ns} generation.
For a light source, we used a \SI{100}{ mW}, \SI{520} {nm} laser light generated by a WSLD-520-001-2 laser diode (Wavespectrum). The laser diode was driven with a home-built diode driver, which stabilized the output power when the laser was on and allowed 100\% on-off modulation of the output power with \SI{<5}{us} rise/fall times. The laser light was focused into a \SI{6 \mathrm{x}11}{ \mu m}. As the sample for our research, we used a diamond plate polished perpendicular to the $\langle 111 \rangle$ crystallographic axis (Velman LLC) with approximately \SI{1}{ppm} NV centers. The NV center fluorescence was collected with a parabolic concentrator made of sapphire, similar to in \cite{Wolf2015}. Both the fluorescence and pumping laser intensities were registered by a photodiode (PDB-C609-2) and amplified with a transimpedance amplifier. The analog signal was digitized by an AD7626 analog-to-digital converter directly connected to an FPGA.

A constant magnetic field was formed by rectangular Helmholtz coils of size \SI{7\mathrm{x}7}{cm} and applied along the $\langle 111 \rangle$  crystallographic axis.
The MW field was formed by an antenna composed of two 5-mm-diameter coaxial loops made of 1 mm copper wire and separated by 4 mm with a diamond sample in between \cite{Soshenko2018}. The antenna was excited with a weakly inductively coupled loop. The signal for the microwave antenna was sourced from two generators (SG384 and SMA100A), frequency modulated by digital-to-analog converters, and controlled directly by an FPGA. On-off gating and additional amplitude modulation were performed via switching circuity based on ZASWA-50-DR- switches. The microwave signal from the switching circuity was amplified by a Minicircuits ZHL-16W-43X+ amplifier and fed via a circulator to the antenna.
The radiofrequency signal was generated by two generators (Agilent A4400 and DS345), selected and switched by a ZASWA-50-DR- circuit and amplified by a VectaWave VBA100-30 amplifier. The amplifier was connected to the antenna, which was formed by $2 \mathrm{x} 10$ loops of \SI{0.2}{mm} copper wire. 
All generators, the MEMS gyroscope and the FPGA board were connected to a laptop, which retransmitted data and control signals through Wi-Fi to the standalone PC that ran control software.
The whole setup was mounted on a 360 degree high load turntable. The turntable was driven by a commutator motor with a maximum speed of \SI{1/3}{Hz}  and manual direction and speed control. 
The whole setup was mounted on a 360 degree high load turntable. Turntable is driven by collector motor with maximum speed of \SI{1/3}{Hz} and manual direction and speed control. 
All the appliancies were powered by UPS (N-POWER Smart-Vision S3000), retaining energy, enough for 15 minute experiment run.

\bibliographystyle{ieeetr}
\bibliography{references}